\documentclass[english,aps,pre,reprint,superscriptaddress, floatfix]{revtex4-1}

\usepackage{babel}
\usepackage{graphicx}
\usepackage{mathrsfs}
\usepackage{amsmath}
\usepackage{amssymb}
\usepackage{hyperref}

\makeatletter
\renewcommand{\vec}{%
  \mathpalette {\overarrow@\vectfill@}}
\def\vectfill@{\arrowfill@~~{\raisebox{-5.5pt}[\p@][\p@]{$\mathord\mathchar"017E$}}}

\begin{document}
\title{Cavity approach for modeling and fitting polymer stretching}
\date{\today}
\author{Francesco Alessandro Massucci}
\email{francesco.massucci@urv.cat}
\affiliation{Departament d'Enginyeria
Qu\'imica, Universitat Rovira i Virgili, 43007 Tarragona, Spain
}
\author{Isaac P\'erez Castillo}
\affiliation{Department of Mathematics, King's College London, London WC2R 2LS, UK}
\affiliation{Instituto de F\'isica, Universidad Nacional Aut\'onoma de M\'exico, P.O. Box 20-364, M\'exico DF 01000, M\'exico}
\author{Conrad J. P\'erez Vicente}
\affiliation{Departament de F\'isica Fonamental, Universitat de Barcelona, 08028 Barcelona, Spain}

\begin{abstract}%
The mechanical properties of molecules are today captured by single molecule manipulation experiments, so that polymer features are tested at a nanometric scale. Yet 
devising mathematical models 
to get further insight beyond the commonly studied force--elongation relation is typically hard. Here we draw from techniques developed in the context of disordered systems to solve models for single and double--stranded DNA stretching in the limit of a long polymeric chain. 
Since we directly derive the marginals for the molecule local orientation, our approach allows 
us to readily calculate the experimental elongation as well as other observables at wish. As an example, we evaluate the correlation length as a function of the stretching force. Furthermore, we are able to fit successfully our solution to real experimental data. Although the model is admittedly phenomenological, our findings are very sound. For single--stranded DNA our solution yields the correct (monomer) scale and, yet more importantly, the right persistence length of the molecule. In the double--stranded case, our model reproduces the well-known overstretching transition and correctly captures the ratio between native DNA and overstretched DNA. Also in this case the model yields a persistence length in good agreement with consensus, and it gives interesting insights into the bending stiffness of the native and overstretched molecule, respectively.
\end{abstract}%

\maketitle
\section{Introduction}
\label{intro}
The development of techniques such as atomic force spectroscopy \cite{Binning1985,Binning1986}, magnetic tweezers \cite{Gosse2002}, or laser optical tweezers \cite{Ashkin1970a,Ashkin1970b} allows us today to manipulate single molecules in the laboratory. These experiments are aimed at unveiling the mechanical properties that are key to the biological processes such molecules are involved with. Understanding the energy scales involved in pulling, bending or twisting a molecule, or the structural conformation this assumes, is fundamental to fully appreciate and possibly control processes that such as DNA replication or RNA translation. In this work we focus in particular on DNA stretching experiments, which have been carried out since the early 1990s \cite{Smith1992,Bustamante1994} and that allowed us, in these two decades, to understand several properties of this molecule. Single DNA manipulation revealed, for instance, that DNA shows an elastic response to small pulling forces and a resistance to bending \cite{Hagerman1988,Bustamante1994}, implying that DNA develops a non negligible correlation along its molecular chain. Furthermore, stretching experiments allowed to uncover the overstretching transition of double--stranded DNA (dsDNA), which consists of a 1.7--fold sudden elongation of the B-DNA ({\em i.e.} the DNA in its well-known helical configuration) when it is stretched by  forces of around 65 pN \cite{Smith1996,Cluzel1996,Williams2002,vanMameren2009,Lebrun1996,Cocco2004,Whitelam2010,Fu2010}.

To go beyond these results and understand them on a more quantitative basis, it is necessary to complement the experiments with some theoretical models. Such models should be, in principle, simple enough to be easily fit to the experimental data in a laboratory, but yet realistic enough to give insights into the actual mechanisms behind the observations. Among the main models in the literature, the freely jointed chain (FJC) \cite{Flory1953} is too simplistic and only works in a limited range of forces \cite{Smith1992, Marko1995}, while the worm like chain (WLC) is not analytically solvable \cite{Fixman1973,Bustamante1994,Marko1995,Samuel2002}. 
Additionally, the FJC does account for the monomer--monomer interaction that makes the molecule stiff, while the WLC completely discards the discreteness one expects to observe in real molecules. The Kratky-Porod (KP) model \cite{Kratky1949}, which has been lately revisited by some more recent works, tries to capture these two aspects together \cite{Storm2003,Chakrabarti2005,Chakrabarti2006,Rahi2008}. 
These models are in turn generally solved with transfer matrices, which are diagonalised by some (variational) eigenvector; an approach that is also adopted to solve the WLC model \cite{Marko1995, Bouchiat1999}. Such an eigenvector, however, has no clear interpretation in terms of the physics of the stretching process \cite{Storm2003, Rahi2008, Chakrabarti2006}, so that gaining insight beyond the usual force--elongation relations is rather involved.

Here we propose a full solution of a KP--like model that relies on the so--called cavity method \cite{Mezard2001}, which, conveniently, is exact on a chain system. Such method is a common technique in the field of disordered systems \cite{Mulet2002, Mezard2002} and has already proved useful in tackling other types of biological problems \cite{Braunstein2008, Weigt2009, Massucci2013}. The equations we derive only assume the polymer to be sufficiently long and the solution we propose is thus as correct as a full numerical diagonalisation of the transfer matrix. The method we present, though, can give more insight into the processes happening locally on the molecule, because the set of equations that lie at the core of our approach allows us to compute, with no extra cost, several observables and molecular properties (such as the correlation length for varying force) when fitting the model to the experimental data. Finally, because we avoid choosing any arbitrary analytical form to carry out the solution, the quantitative outcome of our fits may not be biased by such a choice.

The remainder of this article is thus organised as follows: in the next section we introduce the model and the method adopted to solve it in the case of single--stranded DNA (ssDNA) stretching. In Sec. \ref{sec:dsDNA} we generalise the discussion to dsDNA and its overstretching transition. In both cases we successfully fit our analytical solution to experimental measurements and uncover insightful results. 

\section{Single--stranded DNA} \label{sec:ssDNA}
The KP model describes a sequence of $N$ interacting segments of length $b_B$ stretched by an external force $\vec{f}$. It can be expressed by the Hamiltonian:
\begin{equation} \label{eq:ssDNAHam}
-\mathscr{H}(\hat{\boldsymbol{t}}; \vec{f}) = J_{B}\sum_{i=1}^{N-1}\hat{t}_i\cdot\hat{t}_{i+1} + b_B\sum_{i=1}^N\hat{t}_i\cdot\vec{f}~.
\end{equation}
Here $\hat{t}_i$ is the orientation of monomer $i$, $\hat{\boldsymbol t}=(\hat{t}_1,\ldots,\hat{t}_N)$, and $J_B$ is a ferromagnetic coupling to favour local alignment of the monomers. Such coupling has the dimensions of an energy and fixes the polymer rigidness: the larger $J_B$, the stiffer the molecule. 
In particular, bigger values of $J_B$ yield a greater molecular persistence length $\xi_0$, given by (Appendix \ref{appx:persLength}):
\begin{equation}\label{eq:persistenceL}
\xi_0 = - \frac{b_B}{\log\mathcal{L}(\beta J_B)},
\end{equation}
where $\mathcal{L}(x) = \coth x -1/x$ is the Langevin function. For fixed $b_B$ and $\beta$, the persistence length is a monotonically increasing function of $J_B$. When $J_B$ grows, interactions are stronger and `information' propagates more easily.

The free energy per monomer $\mathscr{F}$ is as usual given by taking the logarithm of the partition function $\mathcal{Z}$:
\begin{equation} \label{eq:ssDNAZ}
\begin{split}
\mathscr{F}&=-\frac{1}{N \beta}\log \mathcal{Z}~,\\
\mathcal{Z}&= \int_{\mathcal{S}^N} {\rm d} \hat{\boldsymbol{ t}}~e^{-\beta \mathscr{H}(\hat{\bf t}; \vec{f})}~,
\end{split}
\end{equation}
where $\mathcal{S}$ denotes the integration over the 3D unit sphere. The elongation per monomer $L(f)$ can be derived by differentiating $\mathscr{F}$ with respect to the force. In doing so it is convenient to choose a reference frame with the $\hat{z}$ axis along the force direction and to study the elongation along $\hat{z}$:
\begin{equation} \label{eq:DPCL}
L(f) := -\frac{\partial \mathscr{F}}{\partial f}=\frac{b_B}{\mathcal{Z}N}\sum_{i=1}^N\int_\mathcal{S} {\rm d} \hat{\bf t}~\left(e^{-\beta\mathscr{H}}\hat{t}_i\cdot\hat{z}\right).
\end{equation}
The elongation at small and large stretching force can be computed by expanding $\mathscr{F}$ in $f$ and $J_B$, respectively (see Fig \ref{fig:example}). One finds that at small $f$ the elongation goes linearly with the stretching force as (Appendix \ref{appx:smallF}):
\begin{equation} \label{eq:smallFL}
L(f) = \beta b_B^2 \zeta_0 f,
\end{equation}
with $\zeta_0$ given by Eq. \eqref{eq:HeisSmallFL}. Formula \eqref{eq:smallFL} reproduces the known elastic properties of DNA at small force \cite{Bustamante1994} and correctly yields the FJC and the WLC results \cite{Flory1953, Marko1995} when performing the $J_B\to0$ and $b_B\to0$ limits, respectively (Appendix \ref{appx:smallF}). Indeed, normalising the elongation as $\ell \equiv L/b_B$, for vanishing $J_B$ Eq. \eqref{eq:smallFL} gives the small force FJC limit $\ell_{\rm FJC} = (\beta b_B/3)f$ \cite{Flory1953}. Conversely, when $b_B\to0$, Eq. \eqref{eq:smallFL} yields the WLC elongation $\ell_{WLC} = (2\beta \xi_0/3) f$,  with $\xi_0$ being the persistence length, Eq. \eqref{eq:persistenceL} \cite{Marko1995}. Note that the $b_B\to 0$ limit is in practice achieved as long as $fb_B\lesssim J_B$, so that for sufficiently small forces (or $b_B$) the model reproduces the WLC elongation (more on this in Sec. \ref{sec:ssDNAfit}).
When $f$ is large, the elongation saturates to $b_B$ (see Appendix \ref{appx:largeF}) and its expansion, reported in Eq. \eqref{eq:largeFL}, reproduces the FJC elongation by evaluating the $J_B\to0$ limit.

For the intermediate range of forces, the model can be solved in a few different ways: one could use for instance transfer matrices \cite{Storm2003,Chakrabarti2005}, noticing that the partition function $\mathcal{Z}$ can be represented as the trace over the $N$--th power of an integral operator, or numerically, by using a Monte Carlo method to minimize the energy \cite{Chakrabarti2006}. Nevertheless, both approaches involve some approximation --analytical, in the first case, and stochastic, in the second-- of the full analytical solution and make the fitting of the model to real measurements (which after all is the main point of building a model of this sort) unnecessarily cumbersome. We thus choose to follow another path: we borrow a standard technique from the field of disordered systems, {\em i.e.} the cavity method \cite{Mezard2001}, and solve the model with the sole assumption of a large polymer limit \footnote{Note, however, that such assumption can be relaxed without changing the solution framework}.
\subsection{Solution with the cavity method}\label{sec:ssDNAcavity}
\begin{figure*}[t!]
\centering
\includegraphics[width=.6\textwidth]{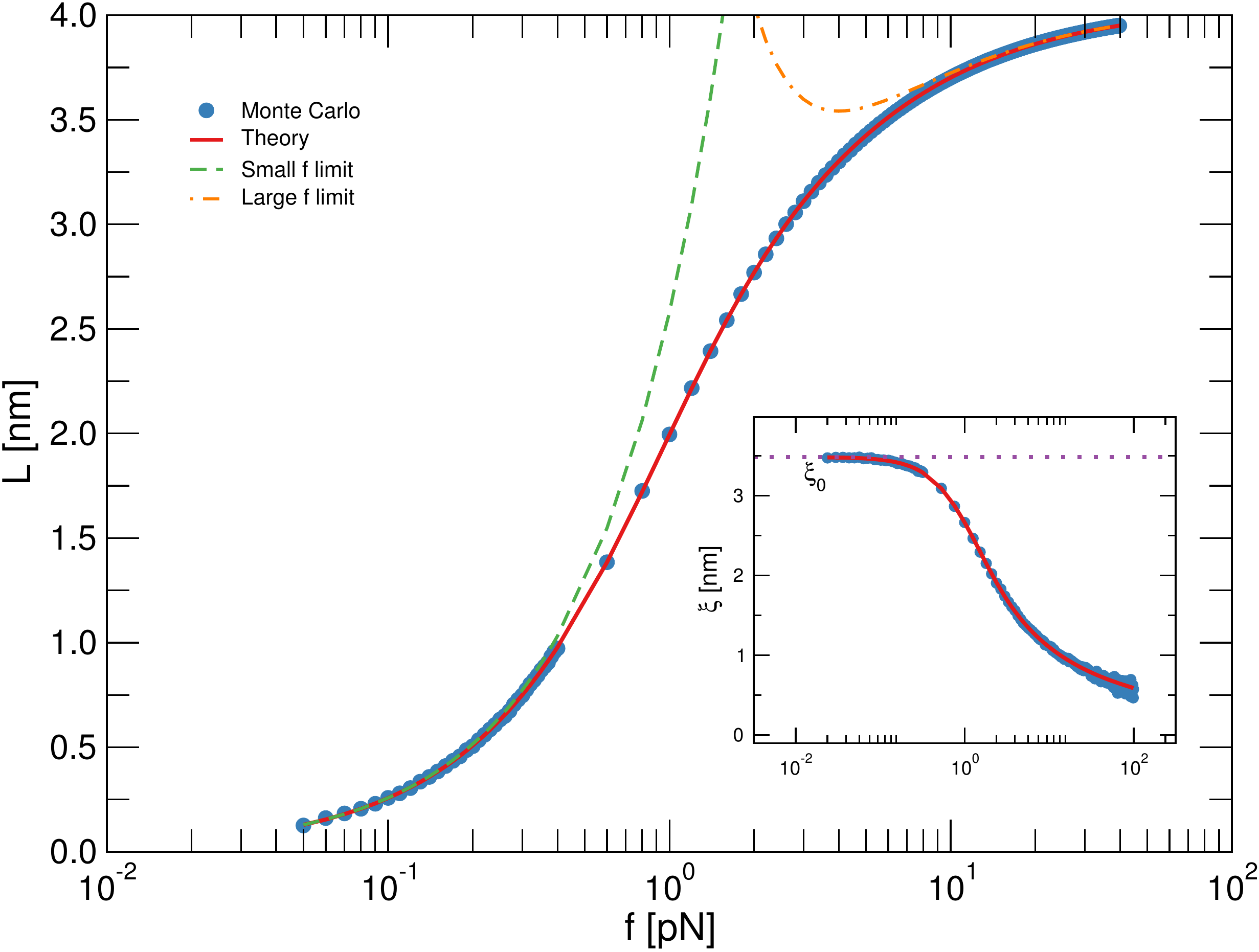}
\caption{(Color online) The cavity method solution {\em vs} heat bath Monte Carlo and large-- and small--$f$ limits. As an example, we fix here $b_B = 4.04$ nm, $J_B=4.04$ nm pN and T = 20C for the value of $\beta$. The main panel shows  the force--elongation curve. Our theoretical solution (red solid line) agrees very well with the Monte Carlo elongation (blue filled circles). Both results, in turn, are also captured by the small (green dashed line) and large (orange dot-dashed line) force limits when $f$ goes to zero or to large values, respectively. Such results confirm the cavity approach may be profitably exploited to study the stretching problem. The inset shows the correlation length $\xi(f)$ as a function of the stretching force. Again the cavity result [Eq. \eqref{eq:persistenceLallF}, red solid line] reproduces extremely well the Monte Carlo simulations (blue filled circles), ensuring that our approach may also be used to evaluate the correlation length for each value of the stretching force. Note that both results correctly converge to the persistence length $\xi_0$, Eq. \eqref{eq:persistenceL}, when $f\to 0$.}
\label{fig:example}
\end{figure*}
We start by noticing that the Hamiltonian \eqref{eq:ssDNAHam} only involves local terms and that the elongation $L$ can be calculated by means of local probability marginals that  factorize. To see this, let us focus for instance on site $i$ in the bulk of the chain. The orientation of such monomer has the marginal probability density function (pdf)
\begin{equation} \label{eq:marginal}
\mathcal{P}_i(\hat{t}_i) = \frac{1}{\mathcal{Z}} \int_\mathcal{S} d{\hat{\boldsymbol{t}}}_{\setminus i} e^{-\beta \mathscr{H}(\hat{\bf t}; f)},
\end{equation}
which is the trace of the Boltzmann weight over all variables except $\hat{t}_i$. Now the Hamiltonian can be rewritten as $\mathscr{H} = -J_B \hat{t}_i \cdot (\hat{t}_{_{i-1}}\hskip -1ex+\hat{t}_{_{i+1}}) - b_B\vec{f}\cdot \hat{t}_i + \mathscr{H}^{(i)}$, where the term $\mathscr{H}^{(i)}$ does not depend on $i$. $\mathscr{H}^{(i)}$ can be further divided into two independent terms, one containing site $i-1$ and its side of the chain, and the second including $i+1$ and the rest. This trick allows to integrate $\mathscr{H}^{(i)}$ over all variables except $\hat{t}_{_{i-1}}, \hat{t}_{_{i+1}}$ and to rewrite Eq. \eqref{eq:marginal}  in the factorized form:
\begin{equation} \label{eq:factMarginal}
\mathcal{P}_i(\hat{t}_i) = \frac{e^{\beta fb_B \hat{t}_i\cdot\hat{z}}}{Z_i}\prod_{j=i\pm1}  \int_\mathcal{S} {\rm d}\hat{t}_{j}~e^{ \beta J_B \hat{t}_i \cdot \hat{t}_j} \mathcal{P}_j^{(i)}(\hat{t}_j),
\end{equation}
where $\mathcal{P}_j^{(i)}(\hat{t}_j)$ is the marginal pdf of site $j$ in the system $\mathscr{H}^{(i)}$ (the so called {\em cavity marginal}) and $Z_i$ is a normalising constant (different from $\mathcal{Z}$). Note that $\mathscr{H}^{(i)}$ is just the system $\mathscr{H}$ where $i$ has been removed. As mentioned, in such system, the two sides of the chain at the left and the right of $i$ are independent and removing $i+1$ does not affect site $i-1$. One can now define $\mathscr{H}^{(i+1)}$ by connecting back site $i$ to $i-1$ and removing $i+1$, so that the pdf of site $i$ reads:
\begin{equation}\label{eq:cavityMarginal}
\mathcal{P}_i^{(i+1)}(\hat{t}_i) = \frac{e^{\beta fb_B\hat{t}_i\cdot \hat{z}}}{Z_i^{(i+1)}}\int_\mathcal{S}{\rm d}\hat{t}_{i-1}~ e^{ J_B \hat{t}_i \cdot \hat{t}_{_{i-1}} }\mathcal{P}_{i-1}^{(i)}(\hat{t}_{_{i-1}}),
\end{equation}
with $Z_i^{(i+1)}$ being a normalising constant. Equation \eqref{eq:cavityMarginal} allows thus to mutually relate the cavity marginals and can be solved iteratively, for each pair of neighboring sites. The fixed point solution of Eq. \eqref{eq:cavityMarginal} can be in turn plugged into Eq. \eqref{eq:factMarginal} to compute the actual marginal of site $i$.

The crucial points of the method are that, since the system is a chain, the factorization performed in Eq. \eqref{eq:factMarginal} is exact and that the same cavity marginals appear in the integral of both Eqs. \eqref{eq:factMarginal} and \eqref{eq:cavityMarginal}.

Note that so far no approximation has been made: in principle, we can choose a suitable chain size $N$, fix appropriate boundary conditions and solve the resulting 2$N$ equations for the cavity marginals iteratively. This would yield to single site marginals that are, by all means, exact. Nevertheless, if we are interested in the properties of the polymer's bulk, it is possible to simplify the system by neglecting boundary effects. We can assume that the molecule is sufficiently long and that, because of homogeneity, each cavity marginal may be written as:
\begin{equation} \label{eq:bulkCavityMarginal}
\mathcal{P}_{\rm cav}(\hat{t}) =\frac{1}{Z_{\rm cav}} e^{\beta f b_B\hat{t}\cdot\hat{z}}\int_\mathcal{S} {\rm d}\hat{s}~e^{\beta J_B \hat{t}_i\cdot\hat{s}}\mathcal{P}_{\rm cav}(\hat{s})~,
\end{equation}
with $Z_{\rm cav}$ being a normalisation constant. We thus choose to solve Eq. \eqref{eq:bulkCavityMarginal} iteratively and use its fixed point solution to evaluate the physical local marginals $\mathcal{P}$ in the polymer's bulk as:
\begin{equation} \label{eq:physicalMarginal}
\mathcal{P}(\hat{t}) =\frac{1}{Z} e^{\beta f b_B\hat{t}\cdot\hat{z}}\left[\int_\mathcal{S} {\rm d}\hat{s}~e^{\beta J_B \hat{t}_i\cdot\hat{s}}\mathcal{P}_{\rm cav}(\hat{s})\right]^2~.
\end{equation}
Note that it is also possible to express the free energy of the system in terms of the cavity marginals \cite{Mezard2003}, so that all we need to do to solve the model is to compute the fixed point solution of Eq. \eqref{eq:bulkCavityMarginal}. With a reasonable choice of the integration routine, such fixed point is reached very quickly, so that the method is also convenient because of its reduced computational cost.

The present approach has another interesting feature, since it allows to easily compute averages of single-- and two--site observables. For instance the elongation per site $L(f)$ of the molecule is simply equal to
\begin{equation} \label{eq:ssDNAelong}
L(f) = b_B \int_\mathcal{S} {\rm d}\hat{t}~\Bigl(\mathcal{P}(\hat{t}) \hat{z}\cdot\hat{t}\Bigr),
\end{equation}
which gives the average of the projection of the molecule along the force axis. 
Two--site averages are instead computed by iterating the equations for the derivatives of the cavity marginals. As an example, we can evaluate the correlation length $\xi(f)$ for any value of $f$. To do so, let us first recall that the susceptibility at fixed force $\zeta_f$ is equivalently given by differentiating twice the free energy with respect to $f$ or by summing the correlation function over the distance $r$:
\begin{equation}
\begin{split}
\zeta_f &=-\frac{1}{\beta}\frac{\partial^2 \mathscr{F}}{\partial f^2}\equiv \frac{b_B}{f}\frac{\partial L}{ \partial b_B},\\
\zeta_f &= \zeta(0)+2\sum_{r=1}^\infty \zeta( r),
\end{split}
\end{equation}
respectively. Now, the first relation can be related to the derivative $\partial_{b_B}\mathcal{P}_{\rm cav}(\hat{t})$ (see Sec. \ref{sec:ssDNAfit} and Appendix \ref{appx:InstabProp} for more details), while the second reduces to a geometric series when assuming that the correlation decays exponentially as $\zeta( r) = \zeta( 0) e^{-r/\xi}$. The constant $\zeta(0) := \langle\hat{t}\cdot\hat{z}\hat{t}\cdot\hat{z}\rangle_C^f$ is the connected (self) correlation at fixed $f$. By expanding the geometric series one has in particular
\begin{equation} \label{eq:persistenceLallF}
\xi(f) = \frac{-b_B}{\log\bigl[\bigl( \zeta_f - \zeta(0) \bigr)/\bigl( \zeta_f + \zeta(0)\bigr)\bigr]}.
\end{equation}
Since we can easily evaluate $\zeta(0)$ with marginal \eqref{eq:physicalMarginal} and $\zeta_f$ by iterating the derivative of the cavity marginal, we are able to get $\xi (f)$ for any value of $f$ at practically no extra cost.

As an illustrative example to validate the method, we can compute some of these quantities and check their values against some numerical simulation. The model described by \eqref{eq:ssDNAHam} can indeed be easily simulated with a Monte Carlo Heat Bath algorithm \cite{Miyatake1986}, so that the accuracy of our calculations can be readily tested. In Fig. \ref{fig:example}, we compare our analytical force--elongation curve, Eq. \eqref{eq:ssDNAelong}, with the outcome of the numerical simulations and the large and small force limits of the theory (Eqs. (\ref{eq:smallFL}, \ref{eq:largeFL}), respectively), for a particular choice of the parameters $b_B$ and $J_B$. The cavity solution agrees extremely well with both the numerical simulations and the two limits, which are computed without relying on the cavity formalism. In the inset of Fig. \ref{fig:example} we show, furthermore, the correlation length at fixed force $\xi(f)$ computed with our method and we compare it with the correlation length evaluated with the numerical simulations. Again, the two results are seen to overlap extremely well and to reproduce, for $f\to0$, the correct persistence length at null force $\xi_0$ given by Eq. \eqref{eq:persistenceL}. These results show that the cavity approach can be successfully applied to solve the KP model. Hence, once the properties of our solution have been characterized, we can move forward in illustrating how the cavity approach comes particularly convenient when fitting the theory to the experimental results.

\subsection{Fit to the experiments} \label{sec:ssDNAfit}
\begin{figure*}[t!]
\centering
\includegraphics[width=0.6\textwidth]{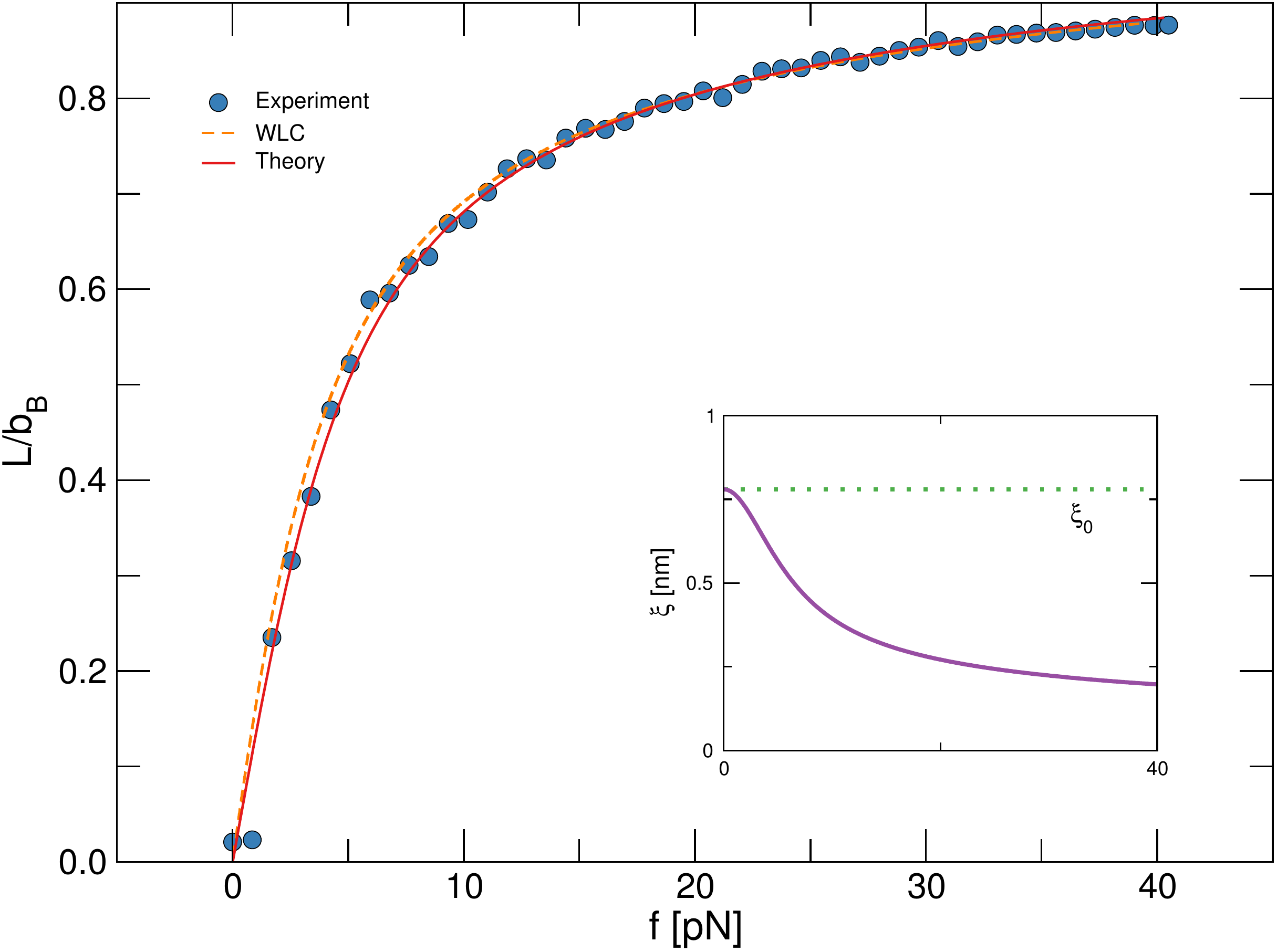}
\caption{(Color online) Model fit to ssDNA stretching experiments. The main panel shows the experimental force elongation (blue circles), {\em vs.}~ the WLC interpolation formula \cite{Marko1995, Hsin2011} (orange dashed line) and the cavity method solution Eq. \eqref{eq:ssDNAelong} (red solid line). The parameter values returned by the fit are $L_{\rm WLC}=2 \mu$m for the WLC contour length and $\xi_0^{\rm WLC}=1.04$nm for the WLC persistence length and $b_B=$0.64 nm, $J_B=$6.13 pN nm for the cavity solution. $\beta$ was fixed assuming room temperature $T=$20C. The measured elongation is normalised over the known number of bases (3000) times the monomer length $b_B$. The agreement between our theory and the measurements is confirmed by the $\chi^2=0.063$, the WLC interpolation fit yields instead $\chi^2=0.098$. The inset shows the correlation length $\xi(f)$ as a function of the stretching force (solid line), computed from Eq. \eqref{eq:persistenceLallF}. It is a decreasing function departing from the value $\xi(0) = 0.78$ nm at zero force (dashed line), {\em i.e.} the persistence length $\xi_0$ of the molecule.
}
\label{fig:ssDNAfit}
\end{figure*}
Our aim is now to find the parameter values $b_B, J_B$ that best fit Eq. \eqref{eq:ssDNAelong} to the experimental elongation. The fit value of $b_B$ and $J_B$ may be used, in turn, to compute other observables such as, for instance, the persistence length, given by Eq. \eqref{eq:persistenceL}, and the correlation length at fixed force, Eq. \eqref{eq:persistenceLallF}. Given the $\mathcal{N}$ experimentally measured force--elongation points $(f^{({\rm exp})}_\kappa,L^{({\rm exp})}_\kappa)$, $\kappa=1, \ldots, \mathcal{N}$ the best guess for the parameter vector $\boldsymbol{\mu} = (J_B, b_B)$ is thus found by least squares minimising the cost function
\begin{equation} \label{eq:chi2}
\chi^2 = \sum_{\kappa=1}^{\mathcal{N}} \left(L^{({\rm exp})}_\kappa - L \left( f^{({\rm exp})}_\kappa ;\boldsymbol{\mu}\right)\right)^2\,,
\end{equation}
where $L ( f^{({\rm exp})}_\kappa ;\boldsymbol{\mu})$ is the value of Eq. \eqref{eq:ssDNAelong} at force $f^{({\rm exp})}_\kappa$ for a fixed parameter vector $\boldsymbol{\mu}=(J_B,b_B)$. Equation \eqref {eq:chi2} measures in practice the distance between the elongation $L$ predicted by our model, Eq. \eqref{eq:ssDNAelong}, and the experimental elongation \footnote{Note that in general the enthalpic elongation of DNA can be accounted for by introducing a stretching modulus $Y$ as in the FJC and WLC \cite{Wang1997}. For us it is not necessary as the experimental values of $f$  are relatively small and the enthalpic elongation is not observed.}. The analytical minimisation of $\chi^2$ is achieved by computing the gradient $\nabla_{\boldsymbol{\mu}} \chi^2$, which ultimately depends on the derivatives of $L$ with respect to the parameters of the model. Interestingly, such derivatives turn out to be correlation functions that may be well studied within the cavity method (see Appendix \ref{appx:InstabProp}). As a consequence, solving the $\chi^2$ minimisation gives additional information on non local processes taking place in the molecule. Direct differentiation of the cavity equations ultimately leads to
\begin{equation} \label{eq:fitCavity}
\begin{split}
\frac{\partial  Z_{\rm c} \left(\hat{t}\right)}{\beta\partial J_B} &= e^{\beta fb_B \hat{t} \cdot \hat{z}}\int_\mathcal{S}{\rm d} \hat{u}\;e^{\beta J_B \hat{t} \cdot \hat{u}}\left[ \hat{t} \cdot \hat{u} Z_{\rm c}  \left(\hat{u}\right)+\frac {\partial Z_{\rm c} \left(\hat{u}\right)}{\beta\partial J_B} \right]\\ 
\frac{\partial Z_{\rm c}\left(\hat{t}\right)}{\beta \partial b_{B}} &= f\hat{t} \cdot \hat{z} Z_{\rm c}\left(\hat{t}\right)+ e^{\beta fb_B \hat{t} \cdot \hat{z}}\int_\mathcal{S}{\rm d} \hat{u} \; e^{\beta J_B \hat{t} \cdot \hat{u}}\frac{\partial Z_{\rm c}\left(\hat{u}\right)}{\beta\partial b_{B}},
\end{split}
\end{equation}
where $Z(\hat{t})=Z\mathcal{P}(\hat{t})$ and its cavity counterpart $Z_{\rm c}(\hat{t})=Z_{\rm cav}\mathcal{P}_{\rm cav}(\hat{t})$. Indeed
\begin{equation} \label{eq:fitdZ}
\frac{\partial \mathcal{P}}{\partial \mu_i}\left(\hat{t}\right)=\frac{1}{Z} \frac{\partial Z\left(\hat{t}\right)}{\partial \mu_i} -\mathcal{P}\left(\hat{t}\right)\frac{1}{Z} \frac{\partial Z}{\partial \mu_i}\,~~~ i=1,2.
\end{equation} 
The fixed point of Eq. \ref{eq:fitCavity} yields $\nabla_{\boldsymbol{\mu}} \chi^2$ and allows to implement the fit.

Therefore, we numerically iterate Eqs. \eqref{eq:bulkCavityMarginal}, \eqref{eq:fitCavity} for $Z_{\rm c}(\hat{t})$, $\partial_{J_{B}} Z_{\rm c}\left(\hat{t}\right)$, $\partial_{b_B} Z_{\rm c}\left(\hat{t}\right)$ respectively. The fixed point solution is then used to compute $\nabla_{\boldsymbol{\mu}} \chi^2$. The value of $\boldsymbol{\mu}$ yielding $\nabla_{\boldsymbol{\mu}} \chi^2=0$ is finally retained as the best fit of the model to the experimental measurements. The key advantage of the approach is that by iterating these few equations, one is able to minimise the $\chi^2$ and to obtain simultaneously the force--elongation relations Eq. \eqref{eq:ssDNAelong}, the correlation length Eq. \eqref{eq:persistenceLallF}, or whichever additional thermodynamic quantity one may be seeking. We stress again that the computational cost to reach the fixed point solution of our equations is very limited: at fixed $f$, convergence is reached on average in 14 iterations. By using Gaussian quadrature integration with 14 weights, which yields a very good estimate of the functions, this is achieved in less then 0.1s on an Intel i7 processor.

We hence go on to fit a stretching experiment performed on a ssDNA chain of 3000 bases in a 10 mM NaCl solution. The interbase distance was estimated to approximately 0.7 nm by x--ray diffraction.  The fit, which is plotted in Fig. \ref{fig:ssDNAfit}, allows us to find a monomer length  $b_B \simeq$ 0.64 nm, {\em i.e.} a value very close to the actual interbase distance. Such finding suggests that the model is able to capture the real molecular fundamental length. We also have for the coupling constant $J_B \simeq 6.13$ pN nm. As mentioned in the previous section (and as explained in Appendix \ref{appx:persLength}), parameters $b_B$ and $J_B$ can be used to compute the persistence length $\xi_0$ of the model. Applying formula \eqref{eq:persistenceL}, we find $\xi_0 \simeq 0.78$ nm, which is also consistent with experimental results \cite{Smith1996,Tinland1997}. 

In Fig. \ref{fig:ssDNAfit} we also show a fit of the WLC interpolation formula \cite{Marko1995, Hsin2011} to the experimental points. The resulting $\chi^2$ is larger than the cavity fit (Fig. \ref{fig:ssDNAfit} caption) and the corresponding persistence length $\xi_0^{\rm WLC}=1.04$ nm is longer than our result, probably due to the approximation made by the interpolation formula  in the intermediate range of forces (see, {\em e.g.}, Fig. 3 in Ref. \cite{Marko1995}). Indeed, fitting the variational WLC formula \cite{Marko1995}, which is a better estimate of the WLC exact solution, yields basically the same curve and roughly the same persistence length of the cavity fit (not shown). This is to be expected, since, as mentioned, our model should reproduce the WLC limit as long as $fb_B\lesssim J_B$, which is the case for a broad range of the experimental forces. Such findings stress thus the importance of choosing an exact approach against an approximated one. In this sense, we feel that our method is a convenient alternative to the WLC and other approaches in the literature \cite{Marko1995, Storm2003}, because, in addition to the persistence length and the molecular fundamental unit $b_B$, it allows us to evaluate other observables at no extra cost. 
In the inset of Fig. \ref{fig:ssDNAfit} we show for instance the correlation length at any force value $\xi(f)$; Eq.  \eqref{eq:persistenceLallF}. This is seen to be a decreasing function of $f$, which correctly equals the persistence length $0.78$ nm at $f=0$.

In conclusion, we have shown in this section how our method allows one to simultaneously compute different observables by simply fitting the theory to the experimental data. We derived indeed very reasonable estimates of the molecule elongation, of the length of its unit blocks $b_B$, of the persistence length $\xi_0$, and of the correlation length at fixed force $\xi(f)$, in a straightforward fashion.

\section{Double--stranded DNA} \label{sec:dsDNA}
Motivated by the accurate results obtained with ssDNA stretching, we now generalise the model to the case of dsDNA and to its structural transition. As explained in the introduction, B-DNA abruptly extends 1.7 times its native length when stretched by $f\simeq65$pN, getting into a state called S-DNA \cite{Smith1996,Cluzel1996}. In the following, we are not making any explicit assumption about the actual nature of such overstretched DNA, but we take into account that it is different, somehow, from the native B-DNA. We thus allow the monomers in Eq. \eqref{eq:ssDNAHam} to be into two possible states, similarly to Refs. \cite{Storm2003,Rahi2008} and in line with the spirit of Ref. \cite{Fiasconaro2012}, where a dynamical model of overstretched DNA featuring a double well potential was studied. For the sake of simplicity we will call the overstretched DNA S-DNA, but we do not make any further speculation on its actual conformation.

To proceed in our modeling, let us reconsider the chain of $N$ monomers described by Eq. \eqref{eq:ssDNAHam}, but let us denote the state of each monomer through a set of binary variables $\boldsymbol{\sigma}=(\sigma_1,\ldots, \sigma_N)$ such that $\sigma_i\in\{B, S\}$, $\forall i=1, \ldots, N$. Since every site $i$ may now be in either $S$ or $B$ state, the parameters $b$ and $J$ must depend on the variables $\boldsymbol{\sigma}$. We allow thus  two different monomer lengths $b_\sigma=(b_B, b_S)$ and three different interaction terms $J_{\sigma\sigma'}=(J_{BB}, J_{BS}, J_{SS})$ (we have $J_{BS}\equiv J_{SB}$).  We introduce some further terms based on energetics considerations: as B-DNA alone is found in nature, we make it energetically more stable than S-DNA by adding an effective field $\gamma_B$ which is different from zero at site $i$ only if $\sigma_i\equiv B$. The parameter $\gamma_B$ can be interpreted as the energy needed to break a B-DNA monomer and transform it into S-DNA, so that increasing it enlarges the force needed to perform the transition. We also include the energetic cost of having a boundary between $S$ and $B$ regions, through the parameter $\epsilon_{BS}$. Varying this parameter changes instead the steepness of the transition. We note in passing that this same setup can be applied, for instance, to study the breaking of secondary structures (helix--coil transition) in protein stretching experiments \cite{Chakrabarti2005,Chakrabarti2006}.

Taking all this into account, our dsDNA Hamiltonian takes the form:
\begin{equation}\label{eq:dsDNAHam}
\begin{split}
\mathscr{H} \left(\hat{\boldsymbol{t}}, \boldsymbol{\sigma};\vec{f} \right) =- \sum_{i=1}^N \Bigl( f b_{\sigma_i} \hat{t}_i\cdot\hat{z}&+\gamma_B \delta_{\sigma_i,B}\Bigr)\\
-\sum_{i=1}^{N-1}  \Bigl( J_{\sigma_i, \sigma_{i+1}} \hat{t}_i \cdot \hat{t}_{i+1}+& \epsilon_{BS}(1-\delta_{\sigma_i, \sigma_{i+1}})\Bigr),
\end{split}
\end{equation}
where $\delta_{ab}$ is a Kronecker delta, equal to one only if $a=b$. It is known  that the fundamental blocs of real DNA are extensible themselves \cite{Smith1992,Marko1995}, and that DNA experiences an enthalpic elongation for large enough stretching forces. To capture this feature we also introduce a {\em stretching modulus} and rewrite the monomer’s length as $b_B = b_B^{(0)} (1 + f/Y )$, with $Y$ the Young modulus \cite{Wang1997} and $b_B^{(0)}$ being the native B monomer length at $f=0$. On the contrary, we assume $b_S$ to be rigid. Considering this, the elongation $L$ along the $\hat{z}$ axis may be again derived by differentiating the free energy with respect to the force modulus $f$:
\begin{equation} \label{eq:dsDNAelong}
L\bigl(f\bigr)= \lim_{N \to \infty}\frac{1}{N}\sum_{i=1}^N\langle\bigl(\delta_{\sigma_{i},B}c_B+\delta_{\sigma_{i},S}b_S\bigr)\hat{t}_i\cdot \hat{z}\rangle.
\end{equation}
Here $\langle\ldots\rangle$ denotes the usual thermal average, while the term $c_B=b_B^{(0)} \left(1+2f/Y\right)$ is produced by differentiating $f b_B(f)$ with respect to $f$.

The elongation and other thermal properties of the system may be again derived comfortably with the cavity method. Similarly to what was done for the ssDNA, one can notice that the Hamiltonian \eqref{eq:dsDNAHam} still depends only on local terms, and, as in Eq. \eqref{eq:factMarginal}, one can write the marginal pdf of each site in terms of the factorized pdf of its neighbours. Discarding boundary effects, one is finally able to express the cavity marginals in the bulk of the molecule as:
\begin{equation} \label{eq:dsDNAcavity}
\begin{split}
\mathcal{P}_{\rm cav} (\hat{t}, &\sigma) = Z_{\rm cav}^{-1}~\exp \Bigl[ \gamma_B\beta\delta_{\sigma,B}+\beta f b_\sigma  \hat{t}\cdot \hat{z} \Bigr]\\
&\times \sum_{\tau\in\{B,S\}} e^{\beta\varepsilon_{\sigma\tau}} \int_\mathcal{S} {\rm d} \hat{u} \; e^{\beta J_{\sigma \tau} \hat{t}\cdot\hat{u} } \mathcal{P}_{\rm cav}(\hat{u},\tau )~,
\end{split}
\end{equation}
where now $\mathcal{P}_{\rm cav}$ depends on the vector $\hat{t}$ and on the binary variable $\sigma$. The constant $Z_{\rm cav}$ assures normalisation so that $\sum_\sigma \int {\rm d}\hat{t}~ \mathcal{P}_{\rm cav} (\hat{t}, \sigma) = 1$. The physical marginal is found by connecting a given site to both its neighbours:
\begin{equation} \label{eq:dsDNAphysMarginal}
\begin{split}
 \mathcal{P} (&\hat{t}, \sigma)= Z^{-1}\exp\Bigl[{\gamma_B\beta\delta_{\sigma,B}+\beta f b_\sigma  \hat{t}\cdot \hat{z}}\Bigr]\\
&  \times \left(\sum_{\tau\in\{B,S\}} e^{\beta\varepsilon_{\sigma\tau}} \int_\mathcal{S} {\rm d} \hat{u} \; e^{\beta J_{\sigma \tau} \hat{t}\cdot\hat{u} } \mathcal{P}_{\rm cav}(\hat{u}, \tau)\right)^2,
 \end{split}
\end{equation}
where $Z$ is again a normalising constant. Equation \eqref{eq:dsDNAphysMarginal} above expresses the probability of finding a monomer in either the $B$ or the $S$ state, oriented along the direction $\hat{t}$. As a consequence, such marginal can in turn be used to compute the elongation as:
\begin{equation} \label{eq:dsDNAcavityElong}
L\left(f\right)=\;\int_\mathcal{S} {\rm d} \hat{t}\,\left(\hat{t} \cdot \hat{z} \right)\left[  c_B\mathcal{P}(\hat{t}, B)+b_{S}  \mathcal{P}(\hat{t}, S)\right]~.
\end{equation}
Furthermore, it is possible to compute the fraction of $B$ and $S$ as the marginal probability of being in state $\sigma=\{B,S\}$ at fixed force:
\begin{equation}
\mathcal{P}(\sigma) = \frac{1}{Z}\int_\mathcal{S}{\rm d}\hat{t} \mathcal{P}(\hat{t},\sigma).
\end{equation}
The cavity approach is thus again particularly convenient: Eq. \eqref{eq:dsDNAcavity} may be solved via iteration and its fixed point solution can be used to compute $L$ and $\mathcal{P}(\sigma)$ by expressing $\mathcal{P}(\hat{t}, \sigma)$ in terms of the cavity marginal. Again, we checked the validity of our approach against Heat Bath Monte Carlo simulations, where we added the $\{B,S\}$ degrees of freedom. Our solution was found to be in excellent agreement with the numerical results.

The strength of the method, though, lies again in that explicit differentiation of the cavity marginals allows one to easily compute the correlations that arise when fitting the model to experimental measurements, as we will show in the next section.
\begin{figure*}
\centering
\includegraphics[width=0.9\textwidth]{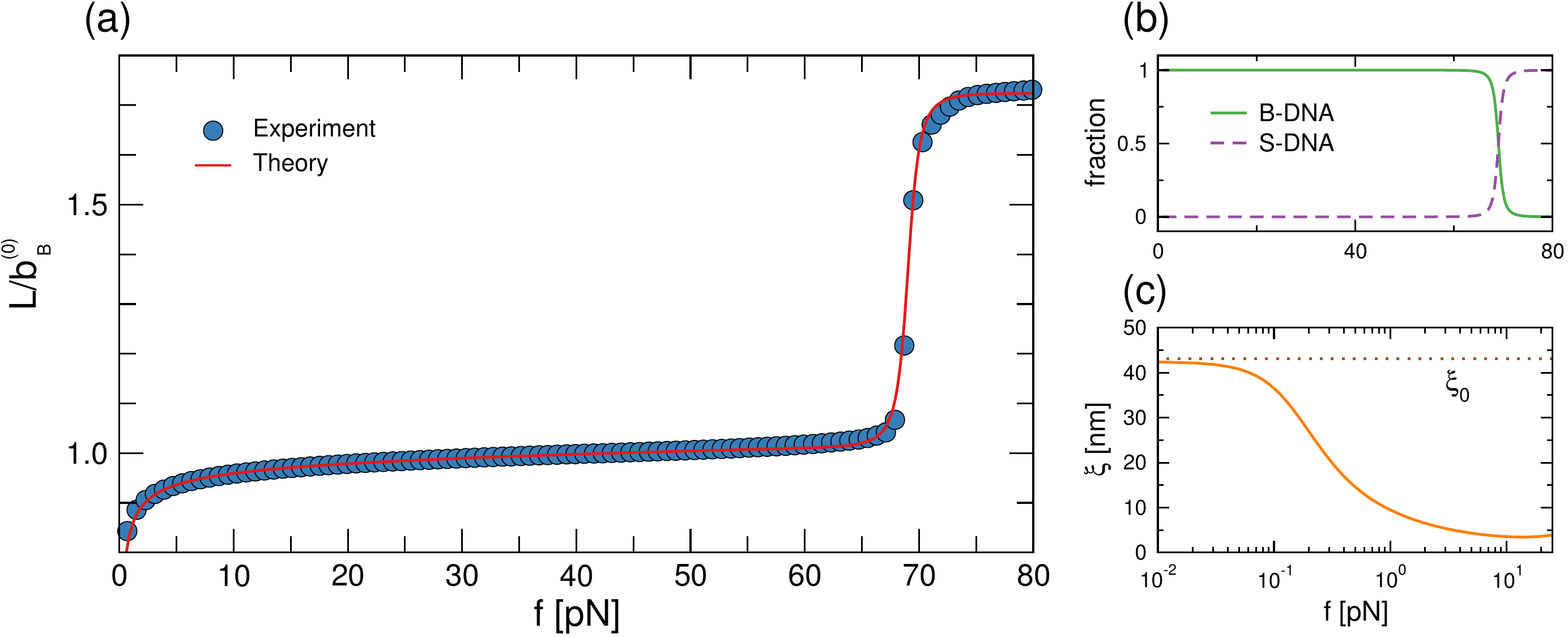}
\caption{(Color online) Model fit to dsDNA stretching experiments. {\bf (a)} The experimental elongation normalised over the native monomer length $b_B^{(0)}$ (blue circles) {\em vs.} the cavity method elongation \eqref{eq:dsDNAcavityElong} (red solid line). To obtain the theoretical curve, we proceeded first to fit the experimental measurements below 50pN with model \eqref{eq:ssDNAHam} and then fixed the values for B-DNA to fit the remaining range of forces with model \eqref{eq:dsDNAHam}. The parameter values yielded by the fit (we we again assumed a temperature T=20C to fix $\beta$) are reported in Table \ref{tab:dsDNAfitRes}. {\bf (b)} the fraction of B-DNA (solid line) and S-DNA (dashed line) as a function of the stretching force $f$. We recall that we do not make any specific assumption on the actual conformation of the S-DNA and that, for us, this is just a generic overstretched phase of B-DNA. We find the overstretching transition to be highly cooperative, in agreement with consensus: S-DNA appears very suddenly around 70pN and it rapidly picks up to the whole molecule. {\bf (c)} The correlation length $\xi(f)$ as a function of the stretching force. It is a decreasing function of $f$ that correctly starts at the persistence length value $\xi_0=43$ nm.}
\label{fig:dsDNAfit}
\end{figure*}

\subsection{Fit to the experimental measurements} \label{ssec:dsDNAfit}
We proceed now to fit model Eq. \eqref{eq:dsDNAHam} to a set of dsDNA stretching measurements. Once again, we fit the theoretical elongation, given by Eq. \eqref{eq:dsDNAcavityElong}, to an experimentally measured one. The parameters to be fit are in this case $\boldsymbol{\mu}=\{$$b^{(0)}_B$, $Y, b_{S}$, $J_B$, $J_{BS}$, $J_\mathcal{S}$, $\gamma_{B}$, $\epsilon_{BS}\}$ and their value is sought again by minimising  a cost function $\chi^2$ of the form of Eq. \eqref{eq:chi2}. We achieve such minimisation by equating to zero the gradient of $\chi^2$ with respect to the vector of parameters $\boldsymbol{\mu}$. The resulting value of the parameters enables us to quantify in turn several properties of the stretched molecule. In essence, $b_B$ and $b_S$ give an estimate of the B and S--DNA monomer lengths respectively, $J_B$, $J_{BS}$ and $J_S$ yield the stiffness of the different portions of the molecule, $\gamma_B$ returns the transition free energy, {\em i.e.} the cost of breaking a B--DNA segment into a S--DNA, while $\varepsilon_{BS}$ is related to the energy penalty of having a B--S boundary. Once more, computation $\nabla_{\boldsymbol{\mu}} \chi^2$ yields a series of correlation functions that may be seen as the propagation of perturbations of the the cavity marginals (see the appendix \ref{appxssec:fit} and \ref{appx:dsDNAcorr}).

Thus, a convenient method to minimise $\chi^2$ is to compute its derivatives by means of the cavity equations. Let us express again the cavity and the physical marginals \eqref{eq:dsDNAcavity} and \eqref{eq:dsDNAphysMarginal} by  $\mathcal{P} (\hat{t}, \sigma)=Z^{-1}Z (\hat{t}, \sigma)$ and $ \mathcal{P}_{_{\rm cav}}(\hat{t}, \sigma)=Z_{\rm c}^{-1}Z_{\rm c} (\hat{t}, \sigma)$, respectively. We have to evaluate the derivatives of $Z (\hat{t}, \sigma)$, which are given by
\begin{equation} \label{eq:dsDNAfitDer}
\begin{split}
\partial_{\mu_i}Z (\hat{t}, \sigma)&=2e^{-\gamma_B\beta\delta_{\sigma,B}-\beta f b_\sigma  \hat{t}\cdot \hat{z}}Z_{\rm c} (\hat{t}, \sigma)\partial_{\mu_i}Z_{\rm c} (\hat{t}, \sigma)\\
&-Z(\hat{t}, \sigma)\partial_{\mu_i}\Bigl(\gamma_B\beta\delta_{\sigma,B}+\beta f b_\sigma  \hat{t}\cdot \hat{z}\Bigr).
\end{split}
\end{equation}
This equation is the key ingredient of the fit; once its fixed point solution is known, one is able to compute the derivatives of $\chi^2$, and by setting them to zero, to minimise it. We thus proceed as before and iterate Eqs. \eqref{eq:dsDNAcavity} and \eqref{eq:dsDNAfitDer} for the cavity marginals and of their derivatives to fixed point. The fixed point solution is used to evaluate $\nabla_{\boldsymbol{\mu}} \chi^2$, which is used to minimize the cost function. The optimal fixed point solution is finally plugged into \eqref{eq:dsDNAphysMarginal}, \eqref{eq:dsDNAcavityElong} to compute the elongation.

Since the model \eqref{eq:dsDNAHam} features several parameters, we choose to divide the fitting procedure into two steps: we use the ssDNA elongation \eqref{eq:ssDNAelong} to fit the experimental measurements below 50 pN, where no overstretching occurs. We then fix the values of $b_B^{(0)}$, $J_{BB}$, and $Y$ to fit Eq. \eqref{eq:dsDNAcavityElong} for larger stretching forces. Our results are summarised in Table \ref{tab:dsDNAfitRes} and plotted in Fig. \ref{fig:dsDNAfit}. 

Let us discuss and interpret the fit results. When fitting Eq. \eqref{eq:ssDNAelong} below 50 pN, we find $b^{(0)}_B=3.22$ nm, {\em i.e.} about 10 base pairs. Such value is interestingly close to the dsDNA helix period (3.3 nm), suggesting that, for B--DNA, one can phenomenologically consider a whole helix twist as a monomer. We also find the Young modulus $Y=4570.38$ pN, which is considerably larger than the typical results obtained with the FJC and the WLC \cite{Storm2003}, implying that our model captures the phenomenology of the stretching up to greater values of the force, without relying on $Y$. We finally uncover $J_B=56.26$ pN nm, which, plugged into Eq. \eqref{eq:persistenceL}, returns a persistence length $\xi_0$ of about 43 nm, which is again in good agreement with dsDNA typical values \cite{Hagerman1988,Marko1995,Bouchiat1999}. Fixing those parameters and fitting Eq. \eqref{eq:dsDNAcavityElong} for forces larger than 50 pN yields the results for the overstretched DNA. We find the unit length $b_S$  to be 5.63 nm, about 1.75 times bigger than the native one, yielding the known ratio between S-DNA and B-DNA total length. Confronting the $J_S$ and $J_{BS}$ values with $J_B$, we see that the native strand is much stiffer (and so more correlated as well) than the overstretched portion of the molecule. This finding may help to understand which type of overstretched DNA we are actually fitting: indeed, as recently shown \cite{Fu2010, King2013}, different experimental conditions in terms of temperature and salt concentration may lead to different overstretched DNA configurations, which can either be melted or double stranded. In our case, the ratio $J_S/J_B\sim 70$ suggests we are actually fitting a melted DNA, because this is expected to be about 60--fold more flexible than B--DNA. The value $\gamma_B=106.61$ pN nm can be used to evaluate the B--S DNA transition free energy per base pair (bp). Since there are about 10 bp per native monomer $b_B$, normalising over this factor such energy is about $2.6\beta^{-1}$, which is again a reasonable estimate of the cost needed to break a B segment into a S--DNA \cite{Cocco2004}. In Fig. \ref{fig:dsDNAfit}b we also show the molecular fraction of B-DNA and (overstretched) S-DNA as a function of $f$. 
We find that, in a very small force range around 70 pN, the molecule goes from an entire B-DNA conformation to a completely overstretched state, implying that the overstretching transition is highly cooperative, in agreement with consensus \cite{Smith1996,Cluzel1996}. The cooperativity is modulated by $\varepsilon_{BS}$, whose value is enough to confer such sharp and narrow shape to the B--S transition. We acknowledge that the resulting $\varepsilon_{BS}=-0.7\beta^{-1}$ value is smaller than what expected for the boundary free energy \cite{Cluzel1996, Cocco2004}, however such energy shall be evaluated from the transition width ({\em e.g.} as outlined in Ref. \cite{Rouzina2001a} for a heteropolymer), rather than from $\varepsilon_{BS}$. Additionally, it must be noted that, in a melting transition as it looks to be our case, the actual cooperativity will ultimately be dominated by the heterogeneity of the dsDNA bps stability \cite{Rouzina2001a}. These effects are not captured by the parameter $\varepsilon_{BS}$ and are beyond the scope of the present model, although they may be experimentally relevant.
Finally, in Fig. \ref{fig:dsDNAfit}c, we plot the correlation length, Eq. \eqref{eq:persistenceLallF}, as a function of $f$. We find it again to decrease when the stretching force gets larger, starting from the correct persistence length value $\xi_0=43$nm when $f=0$.

The results obtained suggest that our model is also able to capture some of the main features of a dsDNA stretching experiment. We are able to fit the experimental data in a quite straightforward manner and to get quick conclusions about the mechanical properties of the molecule. It is perhaps useful to stress that the involved mathematical treatment required by, {\em e.g.}, the transfer matrix method does not allow one to get to the same neat conclusions. The cavity method seems instead a more natural way to solve the problem, thanks to the local quantities it allows to compute, the useful relation between the derivatives of the cavity equations and correlations, and because it yields an exact solution in the case of a long molecule.
\begin{table}
\begin{center}
\begin{tabular}{|c|c|c|c|c|}
\hline
$b_{S}$ & $J_{BS}$ & $J_{SS}$ & $\gamma_{B}$ & $\varepsilon_{BS}$ \\
\hline
5.63 & 26.35 (6.5) & 0.81 (0.2) & 106.61 (2.6) & -2.75 (-0.7)  \\
\hline
\end{tabular}
\caption[Parameters of the fit of the two state DPC model]{Parameters of the fit. The parameter $b_{S}$ is given in nm while $J_{\sigma\sigma'}$ and $\epsilon_{\sigma\sigma'}$  are in pN nm (values in parentheses are in $\beta^{-1}/$bp units, $\gamma_B$ is normalised over the $\sim10$ bp in a $b_B$ segment).}
\label{tab:dsDNAfitRes}
\end{center}
\end{table}
\section{Conclusions}
A paradigmatic issue in modeling polymer stretching is the difficulty in devising a realistic enough yet also mathematically treatable theory. The main models in literature, the FJC and the WLC, both lack some characteristics of real stretched molecules. The variational solution to the KP model, conversely, is an approximation that does not allow the simultaneous evaluation of several observables. In this work we presented an alternative approach based on a KP--like model, which describes the molecule as discrete and stiff at the same time. Our main contribution consists in the full and exact analytical treatment of the problem, through the cavity method, which proves to be very practical to compute different thermal observables and to fit the model to a set of experimental measurements. We were indeed able to comfortably fit two different versions of the model to experimental measurements, one of ssDNA stretching and the other for dsDNA, where the overstretching of the molecule was taken into account.

For ssDNA, we obtained a monomer length of 0.64 nm, a size comparable to the actual interbase distance measured by x--ray diffraction (0.7 nm). We also computed a persistence length of 0.78 nm,  in fair agreement with consensus. The dsDNA model native monomer length was seen to be 3.22 nm, which is intriguingly close to the dsDNA helix period. This result suggests that our model captures as a single monomer a full helix twist. We also found the ovestretched monomer length to be 1.75 longer, reproducing the well known scaling between the two different types of DNA. The B-DNA persistence length was evaluated to be 43 nm, again in good agreement with experimental knowledge.  Also, we estimated the transition free energy per bp to be of the order of $2.6\beta^{-1}$, a value reasonably close to previous estimates. Finally, we saw B-DNA to be far stiffer and to resist more to bending than S-DNA. As recently shown, different overstretched configurations may be obtained depending on the experimental setup: the difference in B-- and S--DNA flexibility we uncovered suggests in our case we dealt with a melted overstretched DNA.

The reasonable results we found encourage us to extend further our approach. Besides the obvious applications in the experimental context, we see some natural extensions of our work. A very similar approach may be used for instance to study molecules subject to random transverse forces \cite{Benetatos2011}, where more than one force is applied to the polymer, or to investigate block copolymers \cite{Blundell2009}. These can be sketched as networks of molecules locally attached together, where one wants to know the force deformation relations when a force is applied. We believe those problems may be profitably studied with a straightforward extension of our approach, thanks to its plain and easy implementation.
\begin{acknowledgements}
We would like to thank Felix Ritort, Joan Camunas and Josep Maria Huguet for providing the experimental measurements carried out at the “Small Biosystems lab” in Barcelona and for discussions.  We kindly thank Roger Guimer\`a for a critical reading of the manuscript and for useful discussions. FAM acknowledges financial support from European Union Grants PIRG-GA-2010-277166 and PIRG-GA-2010-268342. CJPV is supported by the Spanish Mineco through grant FIS2012-38266-C02-02 and by the Generalitat de Catalunya grant 2014-SGR-608.
\end{acknowledgements}
\appendix
\section{The Persistence length} \label{appx:persLength}
We would like here to show how to compute the persistence length, that is, the correlation length at zero force, as a function of the parameters $b_B$ and $J_B$ of the KP model, Eq. \eqref{eq:ssDNAHam}. As explained in Sec. \ref{sec:ssDNA}, $b_B$ fixes the monomers length and $J_B$, which has the dimension of an energy, sets the strength of the monomer interactions along the chain.  When $f=0$, model \eqref{eq:ssDNAHam} is just a Heisenberg chain in zero external field. In this case it is possible to write down the problem in spherical coordinates and to solve it with the transfer matrix method  \cite{Takahashi1999}. In particular, the largest eigenvalue of the matrix gives an estimate of the free energy as $\mathscr{F} = -1/\beta \log \frac{\sinh \beta J_B}{\beta J_B}$, while further expansion in eigenvalues allows to evaluate the correlation function $\zeta (r )$ at distance $r$ as:
\begin{equation} \label{eq:corrFunc}
\zeta(r)=\frac{1}{3}\left(\coth \beta J_B - \frac{1}{\beta J_B}\right)^r~.
\end{equation}
Assuming that the correlation function $\zeta( r )$ decays exponentially, {\em i.e.} $\zeta( r )\sim e^{-r/\xi}$, we can define the persistence (or correlation) length of the model at zero force as written in Eq. \eqref{eq:persistenceL}:
\begin{equation} \nonumber
\xi_0 = - \frac{b_B}{\log\mathcal{L}(\beta J_B)}~,
\end{equation} 
where $\mathcal{L}(x) = \coth x -1/x$ is the Langevin function and $b_B$ is put to fix the length scale.

\section{Small force expansion} \label{appx:smallF}
As shown at the beginning of Sec. \ref{sec:ssDNA}, the elongation per monomer, $L$, of the model is computed by differentiating once the free energy $\mathscr{F}$ with respect to the stretching force $f$. Thus, to study the force elongation relation in the small forces regime, one can perform a Taylor expansion of the free energy about $f=0$ and retain the first terms:
\begin{equation} \label{eq:smallFexp}
\begin{split}
L(f)& = -\lim_{N \to \infty}\frac{1}{N} \sum_{k=0}^\infty \frac{f^k}{k!} \partial_f^{k+1} \mathscr{F} |_{f=0}\\
&\simeq\lim_{N \to \infty}f \frac{\beta b_B^2}{N}\sum_{i,j=1}^N\langle\bigl(\hat{z}\cdot\hat{t}_i\bigr)\bigl(\hat{z}\cdot\hat{t}_j\bigr)\rangle^{^0}_C~,
\end{split}
\end{equation}
where the first term vanishes because of symmetry at $f=0$. Here the subscript $C$ stands for a connected correlation, {\em i.e.} $\langle A B\rangle_C=\langle A B\rangle-\langle A \rangle\langle B\rangle$, while the superscript $0$ denotes that averages are performed over an Hamiltonian with zero external force, {\em i.e.} the well known zero force KP model \cite{Kratky1949} or, equivalently, the one dimensional Heisenberg chain in zero field \cite{Takahashi1999}. Note that such model describes a random walk on a sphere \cite{Ghosh2012}, and can be solved exactly. The sum in Eq. \eqref{eq:smallFexp} is known to give the Heisenberg susceptibility $\zeta_0$ at $f=0$, so that finally the elongation at small $f$ is equal to what shown in Eq. \eqref{eq:smallFL}, {\em i.e.}:
\begin{equation} \nonumber
L (f) \simeq \beta b_B^2 \zeta_0 f,
\end{equation}
where
\begin{equation} \label{eq:HeisSmallFL}
\zeta_0:=\frac{1}{3} \frac{J_B + J_B \coth (\beta J_B)-\beta^{-1}}{J_B - J_B \coth (\beta J_B)+\beta^{-1}}~.
\end{equation}
Formula \eqref{eq:smallFL} reproduces the known elastic DNA response to weak stretching \cite{Bustamante1994} and it also correctly captures the FJC limit when $J_B\to 0$. In such case $\zeta_0 \to 1/3$ and one has, normalising $L$ as $\ell \equiv L/ b_B$:
\begin{equation}
\ell_{\rm FJC} (f) \simeq \frac{\beta b_B} {3}f,
\end{equation}
which is the well known small force limit of the FJC \cite{Flory1953}.

It is also possible to compute the WLC limit of Eq. \eqref{eq:smallFL} by performing $b_B\to0$. To see this, we first note that Eq. \eqref{eq:HeisSmallFL} may be written in terms of the persistence length $\xi_0$, Eq. \eqref{eq:persistenceL}, as:
\begin{equation}\label{eq:HeispersL}
\zeta_0 = \frac{1}{3}\frac{1+e^{-b_B/\xi_0}}{1-e^{-b_B/\xi_0}}.
\end{equation}
Plugging such formula into Eq. \eqref{eq:smallFL} and normalising the elongation, one has
\begin{equation}\label{eq:Lxi}
\ell \simeq\frac{1}{3}\frac{1+e^{-b_B/\xi_0}}{1-e^{-b_B/\xi_0}} \beta b_Bf,
\end{equation}
with $\ell$ once again a normalised elongation ranging from 0 to 1.
The limit $b_B\to 0$ of Eq. \eqref{eq:HeispersL} yields $\zeta_0 \simeq 2\xi_0/(3b_B)$, so that when $b_B\to 0$, Eq. \eqref{eq:Lxi} reproduces the small $f$ WLC elongation \cite{Marko1995}:
\begin{equation}
\ell_{\rm WLC} (f) \simeq\frac{2}{3}\beta \xi_0 f.
\end{equation}

\section{Large force expansion}\label{appx:largeF}
When the stretching force $f$ is much larger than $J_B$,  the contribution of the latter becomes negligible and one is allowed to expand $\mathscr{F}$ in $J_B$ around zero. In such a situation the second term of the Hamiltonian \eqref{eq:ssDNAHam}, which is basically a FJC model, dominates over the first term. Therefore, we expect the large force extension to reproduce the FJC elongation plus corrections. Note, however, that such reasoning is correct as long as the monomer length $b_B$ is finite. If, conversely, one assumes to be in a large force limit but to have $b_B\to0$, one should change the sums to integrals, ending up with the known WLC large force limit, which scales as $\mathcal{O}(f^{-1/2})$.

Keeping a finite length for $b_B$, one may compute again the elongation by differentiation of the free energy as:
\begin{equation} \label{eq:largeFexp}
\begin{split}
&L(f) =- \lim_{N \to \infty}\frac{1}{N} \sum_{k=0}^\infty \frac{J_B^k}{k!} \partial_f\partial_{J_B}^{k} \mathscr{F} |_{J_B=0}\\
&\simeq L_{_{\rm FJC}}(f)+\lim_{N \to \infty}\frac{\beta J_B b_B}{N}\sum_{i,j}\langle\bigl(\hat{z}\cdot\hat{t}_i\bigr)\bigl(\hat{t}_j\cdot\hat{t}_{j+1}\bigr)\rangle^{^{\rm FJC}}_C.
\end{split}
\end{equation}
The first term of the expansion equals to the elongation of the FJC model $L_{_{\rm FJC}}(f)$, given by the Langevin function $\mathcal{L}(x) = \coth x -1/x$. The second term is a connected correlation in the FJC model, which can be evaluated by expressing the scalar products as
\begin{equation*}
\begin{split}
\hat{z}\cdot\hat{t}_i&=\cos \theta_i,\\
\hat{t}_j\cdot\hat{t}_{j+1}&= \sin \theta_j \sin \theta_{j+1} \cos(\phi_j-\phi_{j+1}) + \cos \theta_j \cos\theta_{j+1}~.
\end{split}
\end{equation*}
In the FJC the terms depending on angle $\phi$ vanish when averaging over the whole period, and one is left with the computation of $\sum_{ij}\langle \cos\theta_i \cos \theta_j\cos \theta_{j+1}\rangle_{_{\rm FJC}}^C$. This can be separated into a term having $i=j,j+1$ and another with the remainder. The latter however cancels out when computing connected correlations, as in the FJC all sites are independent. It only remains to evaluate $\sum_j\sum_{i=j,j+1}\langle\cos \theta_i \cos \theta_j \cos\theta_{j+1}\rangle_{_{\rm FJC}}^C$, which, in the thermodynamic limit and for the polymer's bulk, may be cast as:
\begin{equation}
\begin{split}
2 \langle \cos \theta\rangle_{_{\rm FJC}}&\Bigl(\langle \cos^2\theta\rangle_{_{\rm FJC}}-\langle \cos \theta\rangle_{_{\rm FJC}}^2\Bigr) =\\
&2 \mathcal{L}\bigl(\beta b_B f\bigr)\Biggl(1 + \frac{1}{\bigl(\beta b_B f\bigr)^2}-\coth^2\bigl(\beta b_B f\bigr)\Biggr)~,
\end{split}
\end{equation}
where $\mathcal{L}(x)$ is the Langevin function and we  assumed homogeneity dropping indices and replacing the sum with a factor 2. Placing everything together in formula \eqref{eq:largeFexp}, the large force elongation reads:
\begin{equation}
L \simeq b_B  \mathcal{L}(\beta b_B f) \Biggl(1+2 \beta  J_B\Bigl(1 + \frac{1}{\bigl(\beta b_B f\bigr)^2}-\coth^2\bigl(\beta b_B f\bigr)\Bigr)\Biggr).
\end{equation}
Expanding further the Langevin function and the hyperbolic cotangent  for large $f$, one finally gets to:
\begin{equation} \label{eq:largeFL}
L\simeq b_B\Bigl( 1 - \frac{1}{\beta b_B f} + \frac{2\beta J_B}{\bigl(\beta b_B f\bigr)^2}\Bigr)~.
\end{equation}
which predicts saturation when $f\gg 1$. We see that the model adds a correction $\mathcal{O}(f^{-2})$ to the FJC large--force limit, such a correction incorporates indeed the monomers interaction. However, we remark again that the above limit is correct when the monomer length $b_B$ is finite.

\section{Propagating the instability of the cavity equations}\label{appx:InstabProp}
\subsection{Expansion in $f$}\label{appxssec:fExp}
Equation \eqref{eq:smallFexp} relates the small--$f$ elongation to a correlation, which can be also evaluated by direct differentiation in the cavity theory. Similarly to the correlation obtained in \eqref{eq:smallFexp}, differentiation of \eqref{eq:ssDNAelong} yields the following relation in the thermodynamic limit:
\begin{equation} \label{eq:TaylorL}
\lim_{N \to \infty}\frac{1}{N}\sum_{i<j}^N \langle\bigl(\hat{z}\cdot\hat{t}_i\bigr)\bigl(\hat{z}\cdot\hat{t}_j\bigr)\rangle^0_C=  \frac{1}{\beta b_B^2}\frac{\partial L}{\partial f}\Bigr|_{f=0}
\end{equation}
The correlation can be expressed in terms of the distance $k$ between $i$ and $j$
\begin{equation} \label{eq:DfAsCorrelation}
\lim_{N \to \infty}\frac{1}{N}\sum_{i<j}^N \langle\bigl(\hat{z}\cdot\hat{t}_i\bigr)\bigl(\hat{z}\cdot\hat{t}_j\bigr)\rangle^0_C \equiv  \langle\bigl(\hat{z}\cdot\hat{t}_0\bigr)^2\rangle^{^0}_C + \sum_{k=1}^\infty \langle\bigl(\hat{z}\cdot\hat{t}_0\bigr)\bigl(\hat{z}\cdot\hat{t}_k\bigr)\rangle^{^0}_C ~.
\end{equation}
In the cavity theory, Eq. \eqref{eq:physicalMarginal}, one has that $\partial_f\mathcal{P}(\hat{t})$ is produced by differentiating with respect to $f$ the cavity marginal $\mathcal{P}_{\rm cav}(\hat{t})$:
\begin{equation}\label{eq:derL}
\frac{1}{\beta b_B^2}\frac{\partial L}{\partial f}\Bigr|_{f=0}= \int_\mathcal{S}{\rm d}\hat{t}~(\hat{z}\cdot\hat{t})^2\mathcal{P}(\hat{t})\Bigr|_{f=0}+\frac{1}{\beta b_B}\int_\mathcal{S}{\rm d}\hat{t}~\hat{z}\cdot\hat{t}~\mathcal{Q}(\hat{t})
\end{equation}
with
\begin{equation}\nonumber
\mathcal{Q}(\hat{t}) = \frac{2}{Z}e^{\beta f b_B \hat{t}\cdot\hat{z}}\int_\mathcal{S}{\rm d}\hat{s}{\rm d}\hat{u}~e^{\beta J_B\hat{t}\cdot(\hat{u}+\hat{u})}\mathcal{P}_{\rm cav}(\hat{s})\partial_f\mathcal{P}_{\rm cav}(\hat{u}),
\end{equation}
evaluated at $f=0$. Now the first term on the right--hand side of Eq. \eqref{eq:derL} reproduces the local average appearing in the correlation \eqref{eq:DfAsCorrelation}. The remaining terms of relation \eqref{eq:DfAsCorrelation} come from evaluation of $\partial_f\mathcal{P}_{\rm cav}$, {\em viz}
\begin{equation}
\begin{split}
\frac{\partial \mathcal{P}_{\rm cav}}{\partial f}(\hat{t}) = &\mathcal{P}_{\rm cav} (\hat{t})\Bigl(\beta b_B (\hat{z}\cdot \hat{t}) - \frac{1}{Z_{\rm cav}} \frac{\partial Z_{\rm cav}}{\partial f}\Bigr)\\
&+ \frac{e^{\beta tb_B f \hat{z}\cdot \hat{t}}}{Z_{\rm cav}}\int_\mathcal{S} {\rm d} \hat{s}~e^{\beta J_B \hat{t}\cdot\hat{s}} \frac{\partial \mathcal{P}_{\rm cav} }{\partial f}(\hat{s})~.
\end{split}
\end{equation}
Indeed one can write the above expression in a closed form to see the equivalence with the formula \eqref{eq:DfAsCorrelation}:
\begin{equation}\label{eq:closedPropag}
\begin{split}
\frac{\partial \mathcal{P}_{\rm cav}}{\partial f}(\hat{t}_0) &= \beta b_B \sum_{k=0}^\infty Z_{\rm cav}^{-k}\int {\rm d}\hat{t}_k~\mathcal{T}(\hat{t}_0,\hat{t}_k)\mathcal{P}_{\rm cav}(\hat{t}_k) \hat{z}\cdot\hat{t}_{k}\\
\mathcal{T}(\hat{t}_0,\hat{t}_k) &= \mathbb{I}_{k0}\int_{S}\Bigl[\prod_{\ell=1}^{k-1} {\rm d }\hat{t}_{\ell}\Bigr]e^{\beta  \sum_{\ell=0}^{k-1} f b_B \hat{z}\cdot\hat{t}_\ell+ J_B\hat{t}_{\ell }\cdot \hat{t}_{\ell+1}},
\end{split}
\end{equation}
where $\mathbb{I}_{k0}=1+\delta_{k0}\bigl(\delta(\hat{t}_k-\hat{t}_0)-1\bigr)$ is a factor to get properly rid of integration in $\hat{t}_0$. Relation \eqref{eq:closedPropag} allows us to directly relate the correlation between sites $0$ and $k$ to the propagation of perturbations of the cavity marginals from $k$ to $0$. 
 \subsection{Model fit}\label{appxssec:fit}
Similarly to what was done in Sec. \ref{appxssec:fExp}, it is possible to express in closed form Eqs. \eqref{eq:fitCavity} used to fit the model to the experimental data. Explicit differentiation of $Z_{\rm c}(\hat{t})$ leads to an expression where the propagation of the perturbations from sites $k=0, \ldots, \infty$ to a fixed site are clearly seen. Let us consider the perturbation with respect to $b_B$:
\begin{equation}
\frac{\partial Z_{\rm c} }{\partial b_{B}}\left(\hat{t}_0\right)=f\sum_{k=0}^\infty\int_\mathcal{S}{\rm d}\hat{t}_{k} \mathcal{T}(\hat{t}_0,\hat{t}_{k})Z_{\rm c} \left(\hat{t}_{k}\right)\hat{t}_{k} \cdot \hat{z},
\end{equation}
where again integration is not performed for $\mathcal{T}(\hat{t}_0,\hat{t}_0)$. The above expression yields in turn:
\begin{equation}
\frac{\partial Z \left(\hat{t}_0 \right)}{\partial b_B} =f \sum_{k=0}^\infty(2-\delta_{k0}) \int_\mathcal{S}{\rm d}\hat{t}_k~Z(\hat{t}_0,\hat{t}_{k})\left(\hat{t}_{k} \cdot \hat{z} \right),
\end{equation}
where  $Z(\hat{t}_0,\hat{t}_{k})$ is the constrained partition function of a segment of the chain of length $k$:
\begin{eqnarray} \label{DPCderCav1}
Z(\hat{t}_0,\hat{t}_{k})= Z_{\rm c}\left(\hat{t}_0\right)\mathcal{T}(\hat{t}_0,\hat{t}_{k})Z_{\rm c}\left(\hat{t}_k\right).
\end{eqnarray}
It is easy to check that the normalisation of $Z(\hat{t}_0,\hat{t}_{k})$ is precisely $Z$ and we can write $\mathcal{P}(\hat{t}_0,\hat{t}_k) = Z(\hat{t}_0,\hat{t}_{k})/Z$, so that finally one gets to the expression for $\partial_{b_{B}}\mathcal{P}(\hat{t})$:
\begin{equation} \label{eq:dPdb}
\frac{\partial \mathcal{P}}{\partial b_{B}}\left(\hat{t}_i\right)=f \sum_{k=0}^\infty(2-\delta_{k0})\int_\mathcal{S} {\rm d}\hat{t}_k \mathcal{P}(\hat{t}_0,\hat{t}_k)\Bigl[\hat{t}_{k} \cdot \hat{z}  - \langle\hat{t}_{k} \cdot \hat{z}\rangle_{\mathcal{P}}\Bigr],
\end{equation}
where $\langle\ldots\rangle_\mathcal{P}$ is an average evaluated via pdf $\mathcal{P}$. Expression \eqref{eq:dPdb} relates $\partial_{b_B}\mathcal{P}$ to the propagation of a perturbation of the cavity marginal from site $k$ to site $0$ and when plugged into \eqref{eq:ssDNAelong} it yields a sum of correlations between 0 and $k$.

\section{Correlation functions in the dsDNA}\label{appx:dsDNAcorr}
We report here, for the interested reader, the correlations arising when differentiating $L$ \eqref{eq:dsDNAelong} with respect to the parameters of model \eqref{eq:dsDNAHam}. By denoting $\tilde{b}_{\sigma_j}=\delta_{\sigma_j B} c_B + \delta_{\sigma_j S}b_S$, one gets, after some manipulation:
\begin{equation} 
\begin{split}
\frac{\partial L}{\partial b_\sigma}(f) &= \lim_{N \to \infty}\frac{1}{N}\sum_{i=1}^N\Bigl[\langle \tilde{b}_{\sigma_i}/b_{\sigma_i} \bigl(\hat{t}_i \cdot \hat{z} \bigr)\delta_{\sigma \sigma_i}\rangle\\
&\hskip 16ex + \sum_{j=1}^N  \langle\tilde{b}_{\sigma_j}\delta_{\sigma_j \sigma}\left(\hat{t}_i\cdot\hat{z}\right)\left(\hat{t}_j\cdot\hat{z}\right)\rangle_C\Bigr]\\
\frac{\partial L}{\partial J_{\sigma\tau}}(f) &=\lim_{N \to \infty}\frac{1}{N}\sum_{i=1}^{N-1} \sum_{j=1}^N\langle\tilde{b}_{\sigma_j}\delta_{\sigma_i\sigma}\delta_{\sigma_{i+1}\tau}\left(\hat{t}_j\cdot\hat{z}\right)\left(\hat{t}_i\cdot\hat{t}_{i+1}\right)\rangle_C\\ 
\frac{\partial L}{\partial \gamma_\sigma}(f) &=\lim_{N \to \infty} \frac{1}{N} \sum_{i,j=1}^N\langle\tilde{b}_{\sigma_j}\left(\hat{t}_j\cdot\hat{z}\right)\delta_{\sigma_i{B}}\rangle_C\\ 
\frac{\partial L}{\partial \varepsilon_{\sigma\tau}}(f)&=\lim_{N \to \infty}\frac{1}{N}\sum_{i=1}^{N-1} \sum_{j=1}^N \langle\tilde{b}_{\sigma_j}\left(\hat{t}_j\cdot\hat{z}\right)\delta_{\sigma_i \sigma}\delta_{\sigma_{i+1} \tau}\rangle_C~.
\end{split}
\end{equation}
Derivation with respect to $b_B^{(0)}$ and $Y$ follows from $b_B$ through the chain rule.

 \bibliographystyle{apsrev4-1}
 \bibliography{stretching_lite}
%
%
%

\end{document}